%% file: mass.tex
\documentclass[12pt]{article}
\usepackage[margin=1in]{geometry}                 
\usepackage{setspace}
\usepackage{graphicx}
\usepackage{amsmath}
\usepackage{amssymb}
\usepackage{bm}
\usepackage{url}
\usepackage{epstopdf}
\usepackage{times}
\usepackage{courier}
\usepackage{authblk}

\usepackage{algorithmic}
\usepackage{textcomp}
\usepackage{xcolor}
\usepackage{subcaption}
\usepackage{diagbox}
\usepackage{tabulary}
\usepackage{mathtools}
\usepackage{listings, multicol}
\title{MASS: Mobile Autonomous Station Simulation}
\author{Thomas Sandholm and Sayandev Mukherjee}
\affil{Next-Gen Systems, CableLabs}

\begin{document}
\maketitle

\input{abstract}
\input{introduction}
\input{relatedwork}
\input{background}
\input{metrics}
\input{model}
\input{data}

\input{evaluation}

\input{system}
\input{usecase}
\input{discussion}
\input{acknowledgments}

\bibliographystyle{IEEEtran}
\bibliography{related}
\newpage
\appendix
\input{bashoptions}
\newpage
\input{restapi}
\newpage
\input{ns3example}
\newpage
\input{android}

\end{document}

%% file: abstract.tex

\begin{abstract}
We propose a set of tools to replay
wireless network traffic 
traces, while preserving
the privacy of the original traces.  Traces are generated  
by a user- and context-aware trained generative adversarial network (GAN).

The replay allows for realistic traces from any number of users
and of any trace duration to be produced given contextual parameters 
like the type of application and the real-time signal strength.

We demonstrate the usefulness of the tools
in three replay scenarios: Linux-
and Android-station experiments
and NS3 simulations. 

We also evaluate the ability of the GAN model to generate traces that retain key
statistical properties of the original
traces such as feature correlation,
statistical moments, and novelty. Our results
show that we beat both traditional statistical
distribution fitting approaches as well
as a state-of-the-art GAN time series generator
across these metrics. The ability of our GAN model to
generate any number of user traces regardless
of the number of users in the original trace
also makes our tools more practically applicable
compared to previous GAN approaches.

Furthermore, we present a use case where our tools 
were employed in a Wi-Fi research experiment.
\end{abstract}

%% file: introduction.tex
\section{Introduction}\label{sec:introduction}

\subsection{Background}
Network infrastructure design and evaluation often relies on software simulation and hybrid software-hardware approaches using testbeds.  The chief use of these tools is to see how the network responds or supports certain patterns of user behavior.  Naturally, the best way to do so is to collect traces of such user behavior and run them through the network simulator/emulator.  However, this approach is not always feasible, because collecting  representative traces of every possible user behavior is not only very expensive and time-consuming but very likely impossible given the long tail of user behavior even when measured in terms of app usage, for example.  Moreover, privacy concerns may lead to fewer users consenting to the use of their personal usage data in this way.  In practice, there are likely to be relatively few datasets of high-quality curated usage data available for software and hardware tools to employ in order to evaluate network performance.

When validating network infrastructure innovations in simulations 
and testbeds, it is therefore common to use one of two approaches:
(1) replay traces from pre-recorded telemetry of real usage, or (2)
apply statistical models of inter-packet arrival time (IAT), traffic load, etc. The first approach suffers from the following problems: 
\begin{itemize}
\item{privacy liabilities}
\item{large amounts of data needed to simulate many users}
\item{number of users that can be simulated are capped by measurements}
\item{discrepancy between measurement and test environment leading to unrealistic replay load}
\item{trace replay does not react to environment.}
\end{itemize}

It however tends to provide more accurate, and user (reality)-based load dynamics
compared to the second approach of model-based trace generation that is just based 
on distributional properties. Model-based trace generation, however, also 
suffers from the lack of responsiveness to environment changes.
So, the question is how we can combine the best of both worlds and achieve:
\begin{itemize}
\item{no risk of exposure of private data}
\item{realistic trace replay}
\item{deployment with limited data}
\item{ability to scale dynamically to any number of distinct users}
\item{reactivity to the environment.}
\end{itemize}

Furthermore, the solution needs to be easy to deploy in: (1) software simulators such as 
NS3; (2) real hardware such as routers, and (3) mobile phones as a front-ends
to traffic replay tools, such as iPerf.

The approach in the present work proposes an autonomous agent that can both 
\emph{replay} realistic workloads as well as \emph{react} to conditions appearing in the environment.  In order to do either, this agent needs to be able to approximate with high accuracy the probability distribution governing the behavior and usage pattern of arbitrary users of the communication network.  In other words, the autonomous agent must incorporate a generative model. 

The power of deep learning models in various applications has, in recent years, prompted a lot of research into their use for generative models.

\subsection{Deep Generating Models and GANs}
A deep generative model (DGM) is a deep learning model trained to approximate a complicated high-dimensional probability distribution about which too little is known to allow us to define a parametric family of distributions containing it.  The authors of ~\cite{ruthotto2021} call such a distribution \emph{intractable} and provide a unified overview of three kinds of DGMs: normalizing flows (NFs), variational autoencoders (VAEs), and generalized adversarial networks (GANs).  All of these approaches assume that we can approximate the intractable distribution by transforming a known, simpler probability distribution (usually a Gaussian) in a latent space of known dimension.  The choice of this latent space dimension is both difficult and important.

NFs apply only to the small set of problems where the latent space dimension equals the intrinsic dimension of the data.  VAEs overcome this limitation by using a probabilistic model to infer the latent variable, but this inference is not straightforward because the generator is nonlinear.  GANs further avoid the challenges of estimating the latent variable and sample directly from the latent distribution.  However, quantifying the similarity between the generated samples and the training data from the true intractable distribution is highly nontrivial.  The GAN model addresses this by training another deep learning model, namely the discriminator, in tandem with the deep learning generator model.

\subsection{Our proposed GAN model}
In the present work, we propose a generative adversarial network (GAN)~\cite{goodfellow2014} generator based on the C-RNN-GAN work on novel
music score generation in~\cite{mogren2016}. Our approach extends the
generator loss function to fit the data to various statistical properties in
the original trace, adds a conditional gradient-descent optimization heuristic,
and allows for context-dependent training and generation.  Similar time series GANs have been proposed for music score generation in~\cite{yoon2019}. A survey of spatio-temporal GAN research is available in~\cite{gao2020}.

The GAN model (including the generator and discriminator networks) is first trained on traces of user traffic measured in a real system. The generator component of the trained GAN model can then be deployed in any simulation or experimental environment. This not only preserves privacy (because the traces used to train the GAN are never available in the deployment environment), but also reduces the volume of data that 
needs to be shipped for deployment (because the deployed model arrives pre-trained). Training typically requires GPU power but the trained generator model can be run on a CPU. We also expose a REST API for
generating traces more easily from constrained environments, without the need to deploy a full
neural network software stack.

We also note that, to achieve even greater privacy protection and robustness in the generated data, we may train the GAN model using Federated Learning~\cite{kairouz2021} with the local learners (clients) training their copies of the model on locally generated and recorded data.  The advantage of Federated Learning is that this local data is never exchanged with any other learner.  Only GAN model parameter updates during training are sent to a central controller that aggregates them to obtain the overall model parameter updates, which are then broadcast to all local learners.  Note that this aggregation could be done securely in a way that further protects the privacy of the local data at the learners, as described in~\cite{sandholm2021}.  A survey of GANs focusing on privacy and security including under distributed learning is given in~\cite{cai2021}.

In terms of contextual awareness and reaction to changing conditions we are mainly interested in task and RF capacity awareness.
Task awareness is the knowledge of which app is running, as it would impact the network demand. Similarly, RF capacity awareness can be inferred from the signal strength, RSSI or similar measures capping the achievable throughput. 

Our basic approach is to begin by generating different time series of
network usage based on different conditions, i.e. apps running and RSSI values. 
Then we apply task-transition probabilities to
``jump'' between the time series on parallel timelines. In an experiment, these jumps could also be triggered by the experimental conditions,
e.g. when a station moves closer to an access point or base station, it jumps to the time series corresponding to the ``strongest-signal'' timeline.

\subsection{Intended Use and Contribution}
We call our new GAN model and the system to provision it {\it MASS}, Mobile
Autonomous Station Simulation. We anticipate that the set of tools that comprises the
system, as well as the general
approach may be valuable to any entrants into a new market, where they could
have developed some great ML algorithm, but do not have enough data to train their model or fully
verify it to put it into production. One example would be Cable operators
entering the mobile operator market.  We also anticipate that the separation
of model training and trace replay would help individual researchers, and practitioners evaluate 
their innovations without direct access to network operator data. Even for organizations
with access to rich data it would limit the need to use up bandwidth to move around
collected telemetry data that instead could be used for customer traffic.

Our key contribution is three-fold:
\begin{itemize}
\item{First, we propose a novel extension to GAN models to retain statistical properties
from the original data (correlations and moments).}
\item{Second, we implement a mechanism to train and replay traces produced by multiple 
GAN models based on environmental contexts (e.g., RF signal and task).}
\item{Third, we design a system and a set of tools that make it easy to replay traces on demand from experiments
and simulations in a wide range of environments (e.g., Android, NS3, OpenWrt).} 
\end{itemize}

The rest of this paper is organized as follows.
We discuss related work and the foundation behind Generative Adversarial Networks
in Section~\ref{sec:relatedwork} and Section~\ref{sec:background}, respectively.
In Section~\ref{sec:metrics}, we present metrics used to evaluate generated traces.
Section~\ref{sec:model} details our model, including the extensions to pre-existing
GAN models. Then we present the data collected from real systems
and used in our study in Section~\ref{sec:data}.
In Section~\ref{sec:evaluation}, we evaluate the traces our model generates
using the metrics defined. 
Section~\ref{sec:system} gives an overview of the system design and describes 
how the MASS tools we developed can be used in three different settings,
Unix Shell, Android, and NS3. In Section~\ref{sec:usecase} we present a
Wi-Fi experiment use case utilizing our tools, and finally in Section~\ref{sec:discussion} 
we provide concluding remarks.

%% file: relatedwork.tex
\section{Related Work}\label{sec:relatedwork}
GANs were first proposed in~\cite{goodfellow2014}.  As remarked in the Introduction, the GAN uses one deep learning model for the generator (thereby making a GAN a kind of DGM) and another deep learning model (called the discriminator) to quantify the difference or similarity between the samples generated by the generator and the training data samples obtained from the real-world intractable distribution that the generator is trying to approximate.  The generator and discriminator are set up in an \emph{adversarial framework}, thus giving rise to the name GAN for the combination of the generator and discriminator.  

General properties of the adversarial framework in general, and GANs in particular, for implicitly learning probability distributions from a statistical point of view, are discussed and studied in~\cite{liang2021}. A survey of GAN models for generation of spatio temporal data is available in~\cite{gao2020}, where the authors provide examples of GAN models to generate time series, spatio-temporal events, spatio-temporal graphs, and trajectory data.  

For our purposes, the most relevant application is time series generation.  One of the earliest GAN models for time series generation was the so-called continuous RNN GAN (C-RNN-GAN) model proposed in~\cite{mogren2016} for music score generation.  The generator and discriminator of this model are not feedforward but recurrent neural networks (RNNs), hence the name.  Since it can generate continuous-valued sequential data of any kind, it is a good candidate for adapting to generate app usage traces for our scenarios of interest.  As we will describe in Sec.~\ref{sec:model}, the key changes we make to the C-RNN-GAN model are to make it context-aware, and to change the definitions of the loss functions at the generator and discriminator from the conventional definitions used by C-RNN-GAN.

A more recent GAN model for generating time series data is the TimeGAN model proposed in~\cite{yoon2019}. The explicit design goal of TimeGAN is to preserve temporal autocorrelations in the generated sequences.  For this purpose, the authors propose a supervised loss function (in addition to the conventional loss functions) as well as an embedding network to provide a reversible mapping between the latent space and the generated feature space.  The embedding and generator networks are jointly trained to minimize the supervised loss. The authors claim that TimeGAN combines the flexibility of a GAN model with the control over the correlations of the generated sequences that is possible with a classical autoregressive model.  However, we see that in the scenario of interest to us, the requirement that the embedding network implement a \emph{reversible} mapping implies that TimeGAN can only generate traces for the same number of users as in the training data, whereas our adaptation of C-RNN-GAN does not suffer from this limitation.  We evaluate and compare our C-RNN-GAN versus TimeGAN in Sec.~\ref{sec:evaluation}.

Yet another recent GAN model for generating time series is R-GAN~\cite{fekri2020}, proposed for generating electricity consumption traces.  Like C-RNN-GAN, R-GAN also uses RNN architectures in the generator and discriminator networks.  However,~\cite{fekri2020} imposes a pre-processing requirement, namely manual extraction of time-series features from the raw data by first fitting a classical auto-regressive integrated moving average (ARIMA) model to it.  We will not compare our results against~\cite{fekri2020} since we are interested in a machine learning workflow that does not rely on manual feature extraction at any stage.

%% file: background.tex
\section{Generative Adversarial Networks}\label{sec:background}

In this section, we will briefly review the adversarial framework and specifically the GAN architecture before providing an overview of the C-RNN-GAN model that we adapt and employ in the present work.  For purposes of comparison, we will also provide a brief description of the TimeGAN model.

\subsection{Adversarial Framework and GANs}
We shall adopt the formalism of~\cite{liang2021} in this section.  The general formulation of the adversarial framework is as follows: we are given a target probability distribution $\nu$ that we need to approximate with a simulated probability distribution $\mu$ obtained from a class of generator distributions $\mathcal{G}_{\mathrm{dist}}$.  The approximation should be so as to minimize the loss incurred over a family of functions inside a discriminator class $\mathcal{D}$ as follows:
\begin{equation}
	\min_{\mu \in \mathcal{G}_{\mathrm{dist}}} \max_{f \in \mathcal{D}} \mathbb{E}_{Y \sim \mu} f(Y) - \mathbb{E}_{X \sim \nu} f(X) = \min_{\mu \in \mathcal{G}_{\mathrm{dist}}} \max_{f \in \mathcal{D}} \int f(\,\mathrm{d}\mu - \,\mathrm{d}\nu) \stackrel{\mathrm{def}}{=} \min_{\mu \in \mathcal{G}_{\mathrm{dist}}} d_{\mathcal{D}}(\mu, \nu),
	\label{eq:gan_math1}
\end{equation}
where we see that the discriminator class $\mathcal{D}$ of functions induces the so-called integral probability metric (IPM) $d_{\mathcal{D}}(\mu, \nu)$ which quantifies the closeness between the generated distribution $\mu$ and the actual target distribution $\nu$.  Different choices for the function class $\mathcal{D}$ describe the various GAN models in use today, including Wasserstein GAN, Maximum Mean Discrepancy GAN, and Sobolev GAN among others.

In a GAN model, both the generator distribution class $\mathcal{G}_{\mathrm{dist}}$ and the discriminator class $\mathcal{D}$ are parameterized by deep neural networks.  So, instead of working with the discriminator class $\mathcal{D}$ of test functions $f$, we shall work directly with the test functions themselves.  Since these functions are implemented by neural networks, we will refer to a function in $\mathcal{D}$ by $D_\omega$ instead of $f$, where $\omega$ represents the parameters of the neural network architecture that implements $D_\omega$. 

Moreover, $\mathcal{G}_{\mathrm{dist}}$ is the class of implicit distributions realized by neural network transformations of a simple lower-dimensional \emph{latent} random variable, for instance a multi-dimensional Gaussian distribution.  Instead of working with the class of distributions $\mathcal{G}_{\mathrm{dist}}$, we will now work directly with the neural network-implemented functions $G_\theta$ that transform the latent input $Z$ to a generated sample $G_\theta(Z)$, where $\theta$ represents the parameters of the functions of the generator neural network architecture $\mathcal{G}$. The relationship between the generator function class $\mathcal{G}$ and the generator distribution class $\mathcal{G}_{\mathrm{dist}}$ is given by
\[
	\mathcal{G}_{\mathrm{dist}} = \{\text{Distribution of } G(Z): G \in \mathcal{G}\}.
\]

Note that the IPM $d_{\mathcal{D}}(\cdot,\cdot)$ is not known analytically because neither the target distribution $\nu$ nor the generated distribution $\mu$ is known analytically.  However, we have $n$ training samples from the target distribution $\nu$, and we generate, say, $m$ samples from the generator distribution $\mu$.  In other words, the GAN generator network should be trained (i.e., its parameters $\hat{\theta}_{m,n}$ selected) so as to
\begin{equation}
	\min_{\theta:G_\theta \in \mathcal{G}} \max_{\omega:D_\omega \in \mathcal{D}} \left\{\hat{\mathbb{E}}_m \left[D_\omega\left(G_\theta(Z)\right)\right] - \hat{\mathbb{E}}_n[D_\omega(X)]\right\},
	\label{eq:gan_math2}
\end{equation}
where the two expectations in~\eqref{eq:gan_math1} are now replaced by empirical estimates $\hat{\mathbb{E}}_m$ and $\hat{\mathbb{E}}_n$ based on the generated and observed samples respectively.  In place of $d_{\mathcal{D}}(\mu,\nu)$,~\cite{liang2021} derives bounds on the difference between the implicit distribution estimator, i.e., the distribution of $G_{\hat{\theta}_{m,n}}(Z)$, and the target $\nu$ under various metrics.

\subsection{The C-RNN-GAN Model}
\label{sec:crnngan}
The C-RNN-GAN~\cite{mogren2016} employs recurrent neural networks (RNNs) implemented by the Long Short-Term Memory (LSTM) architecture for both the generator and the discriminator, with the additional feature that the discriminator's RNN uses a bidirectional LSTM architecture so as to enhance its ability to discriminate between generated samples and samples from the true distribution.    

Although~\eqref{eq:gan_math2} represents the optimization problem for joint training of the generator network $G_\theta$ and the discriminator network $D_\omega$, these two networks are not trained in practice by directly trying to solve~\eqref{eq:gan_math2}.  Instead, the generator network $G_\theta(\cdot)$ is trained to generate samples that fool the discriminator network $D_\omega(\cdot)$, which in turn is trained to be able to discriminate between generated samples and samples from the real target distribution.  

In the notation of the original GAN model~\cite{goodfellow2014}, the output of the discriminator network is the probability that the input to the discriminator network is classified as a sample coming from the true distribution. Thus, the training objective of the generator network is to maximize this probability for generated samples, or equivalently to minimize the complement of this probability.  Similarly, the training objective of the discriminator network is to maximize the complement of this probability for generated samples while also maximizing this probability for training samples.  Thus, the loss functions for training the generator and discriminator networks can be defined as follows~\cite[Alg.~1]{goodfellow2014}:
\begin{align}
	L_{\mathcal{G}} &\stackrel{\mathrm{def}}{=} \frac{1}{k} \sum_{i=1}^k \log(1 - D_\omega(G_\theta(z_i))), \label{eq:gan_genloss} \\
	L_{\mathcal{D}} &\stackrel{\mathrm{def}}{=} \frac{1}{k} \sum_{i=1}^k \left[-\log D_\omega(t_i) - \log(1 - D_\omega(G_\theta(z_i)))\right], \label{eq:gan_discloss}
\end{align}
where $z_1,\dots,z_k$ represents a minibatch of samples from the latent distribution and $t_1,\dots,t_k$ is a minibatch of training samples from the target distribution.  

The C-RNN-GAN retains the $L_{\mathcal{D}}$ of~\eqref{eq:gan_discloss} but modifies the loss function for the generator to be the sum-squared difference (over the minibatch) between the \emph{representations} of the true samples $t_i$ and the generated samples $G_\theta(z_i)$, where the representations are defined by the \emph{logits} of the discriminator neural network, i.e., the last layer before the softmax output layer of the discriminator~\cite[Sec.~3]{mogren2016} -- see also Sec.~\ref{sec:model}.  This is done in order to induce feature matching and reduce the possibility of overfitting of generator network output to the discriminator network.  As will be seen in Sec.~\ref{sec:model}, in the present work we modify the C-RNN-GAN loss functions for training the generator and discriminator networks in several ways from the choices in~\cite{mogren2016}.

\subsection{The TimeGAN model}
Unlike other GAN models including the C-RNN-GAN model, in the TimeGAN model~\cite{yoon2019} the output of the generator network is in the latent variable space instead of in the simulated sample space.  Mappings from the simulated sample space to the latent space (called embedding) and vice versa are done by two other new neural networks, called the embedding and recovery networks respectively.  Given the recovery network function, the generator and discriminator are trained with the losses~\eqref{eq:gan_genloss} and~\eqref{eq:gan_discloss} respectively.  However, the binary adversarial feedback from the discriminator network may not itself constitute sufficient information for the generator network to capture stepwise conditional distributions in the data.  Thus~\cite{yoon2019} proposes an additional supervised loss function for training the generator function, where the labeled target for the output of the generator is the next latent variable given the previous values of the sequence of latent variables.  The generator and discriminator neural networks are trained on a weighted sum of this supervised training loss and the sum of the usual GAN loss functions~\eqref{eq:gan_genloss} and~\eqref{eq:gan_discloss}, whereas the embedding and recovery neural networks are trained on a weighted sum of this supervised training loss and the mean-squared reconstruction loss in going from the simulated sample space to the latent space and back again.

%% file: metrics.tex
\section{Preliminaries: Evaluation Metrics}\label{sec:metrics}
Recall that in Sec.~\ref{sec:crnngan}, we said that we would need to make considerable modifications to the C-RNN-GAN loss functions for the generator and discriminator networks in order to adapt the C-RNN-GAN architecture for our intended purposes of trace generation.  We will describe in detail our choices for these loss functions in Sec.~\ref{sec:model}.  However, even before designing and evaluating the model,
we need to decide on evaluation metrics that can be used to both validate generated traces during training and benchmark against alternative models.

The details of the Mobile Phone Use Dataset~\cite{pielot2017} of user traces on which we develop and evaluate our model may be found in Sec.~\ref{sec:data}.  For now it suffices to mention that although our general approach is data agnostic,
we have observed from the dataset that there
tends to be a strong correlation between download
and upload volumes in the same time period.
Hence, we cannot simply evaluate how well we 
fit statistical properties of individual
upload and download traces.  Instead, we need to fit
the dynamics between features as well.  

Note that the GAN discriminator loss function~\eqref{eq:gan_discloss} does not explicitly incorporate correlations in generated sample sequences.  Moreover, the discriminator network just acts on single traces as inputs, classifying them as either generated by the generator network, or coming from the true distribution.  In particular, the discriminator network does not have at its input a pair of traces, one generated and one from real data, whose statistics it can compare.  Thus the GAN generator cannot expect to receive any feedback from the discriminator network regarding the match between the statistical properties of the generated traces to the traces from the true distribution.  It follows that the GAN generator can only be made to fit these statistical properties in its generated traces if the generator training loss function incorporates such properties. Below, we describe three metrics that we use to measure the goodness of fit between generated samples and samples from the training data, i.e., the true distribution.  In Sec.~\ref{sec:model}, we show how to define a training loss for the generator network of the GAN using these metrics.

\subsection{Correlation distance metric}
\label{sec:metric_corr_dist}
Our first metric is the \emph{Pearson cross-correlation coefficient}
between features. In contrast to traditional approaches
where correlations are either minimized to pick optimal
predictive features or maximized to find interesting
proxy predictors, here we want to fit the level of
correlation between the original trace and the generated
trace. 

To capture the typical user behavior, we proceed as follows:
\begin{enumerate}
\item For each pair of features (e.g., download and upload), compute the Pearson correlation coefficient of the two sequences of samples corresponding to these two features in the trace data for every user in turn.  Thus, if there are $M$ users whose traces are recorded in the training data, then we have $M$ values of the Pearson correlation coefficient for each pair of features.
\item Compute the mean of the above $M$ values of the Pearson correlation coefficient for every pair of features.  This averaging across users is important since some users may have longer
traces captured than others and the longer traces may dominate
the shorter ones if this averaging is not done.
\item If there are $N$ features, then the values of the Pearson correlation coefficient computed in the above step fill an $N \times N$ symmetric matrix, with $n = N(N-1)/2$ distinct (off-diagonal) entries.  These $n$ distinct entries (each entry being the average across users of a Pearson correlation coefficient between a pair of features) may be obtained by simply extracting the (off-diagonal entries from the) upper-triangular part of the above matrix.
\item Write the above $n$ extracted entries into a vector by following some fixed order.  Denote by $\bm{r}_{\mathrm{data}}$ this $n$-dimensional vector of averaged (across users) Pearson correlation coefficients for all pairs of features (computed from the trace data).
\item Now repeat all the above steps for the set of traces generated for each user by the generator network, and denote by $\bm{r}_{\mathrm{gen}}$ the resulting $n$-dimensional vector of averaged (across users) Pearson correlation coefficients for all pairs of features (computed from the generated traces).
\item The {\it \bf correlation distance} $\rho_{\mathrm{corr-dist}}$ is defined as the Euclidean distance
$\|\bm{r}_{\mathrm{data}} - \bm{r}_{\mathrm{gen}}\|$ between the two vectors of Pearson correlation coefficients:
\begin{equation}
  \rho_{\mathrm{corr-dist}} = \sqrt{\sum_{i=1}^n{(r_{\mathrm{data},i}-r_{\mathrm{gen},i})^2}}, 
  \label{eq:rhocd}
\end{equation}
where
\[
	\bm{r}_{\mathrm{data}} = [r_{\mathrm{data},1},\dots,r_{\mathrm{data},n}], \quad \bm{r}_{\mathrm{gen}} = [r_{\mathrm{gen},1},\dots,r_{\mathrm{gen},n}].
\]
\end{enumerate}

For just $N=2$ features such as download and upload, the vectors $\bm{r}_{\mathrm{data}}$ and $\bm{r}_{\mathrm{gen}}$ reduce to scalars $r_{\mathrm{data}}$ and $r_{\mathrm{gen}}$ respectively, where $r_{\mathrm{data}}$ is the correlation coefficient between
uploads and downloads for the source traces and $r_{\mathrm{gen}}$
is the corresponding statistic for the generated data. In this case, the correlation distance is just
\begin{equation}
	\rho_{\mathrm{corr-dist}} = \sqrt{(r_{\mathrm{data}}-r_{\mathrm{gen}})^2}.
	\label{eq:rhocd2}
\end{equation}
Note that this metric is easily differentiable, which is a requirement for including it
in a loss function for training the generator by gradient descent, as we will see in Sec.~\ref{sec:model}.

\subsection{Moments distance metric}
In addition to a good cross-feature correlation fit, we also
want our generated traces to match distributional
properties of the original traces. 

In the present work, we propose to use metrics that measure the discrepancies between statistical moments computed on the original trace data and on the generated traces.  Our motivation for this choice of metric is partly because it yields easily differentiable metrics, and partly because our experiments show that it suffers less from overfitting than direct measures of distribution fitting like KL divergence.

Our second metric {\it \bf moments distance} is the squared Euclidean distance between two three-dimensional vectors, one vector computed on source trace data, the other computed on generated traces.  Each entry in one of these vectors corresponds to a certain moment computed on the appropriate traces (either source data or generated)\footnote{Users with very short traces may not produce reliable values,
of higher moments in particular, and we hence compute this metric across all user trace
steps lumped together.}.  The moments that define the three entries of each vector are respectively the mean $\mu$ (i.e., the first moment), the standard deviation $\sigma$ (i.e., the square root of the second central moment) and skewness $\tilde{\mu}_3$ (i.e., the third standardized moment) of the sequence of samples (across all users) that constitute the corresponding trace data (source data or generated). 

The moments distance metric $\rho_{\mathrm{mom-dist}}$ is thus given by:
\begin{equation}
  \rho_{\mathrm{mom-dist}} = (\mu_{\mathrm{data}} - \mu_{\mathrm{gen}})^2 + (\sigma_{\mathrm{data}} - \sigma_{\mathrm{gen}})^2 + (\tilde{\mu}_{3,\mathrm{data}} - \tilde{\mu}_{3,\mathrm{gen}})^2.
  \label{eq:rhomd}
\end{equation}
Note that in contrast to $\rho_{\mathrm{corr-dist}}$, the moments distance metric $\rho_{\mathrm{mom-dist}}$ is the square of a Euclidean distance and not a Euclidean distance itself.  We can work with the squared Euclidean distance, which is easier to differentiate, for the moments distance metric because the moments themselves are not normalized. In fact, we observed that the generator models trained using the moments distance metric with normalized moments by the algorithm discussed in Sec.~\ref{sec:model} have worse performance than if we leave the moments un-normalized.

\subsection{Novelty metric}
The final metric we define is the {\it \bf novelty}. It is a measure of the novelty of the set of generated traces across all users. Hence it is not a measurement between the source and generated traces.  Rather, it is an internal cross-user metric computed across a set of traces (one trace per user).

The novelty metric is based on the cross correlations (for a selected feature, say download) between the two sequences of samples corresponding to the traces for a pair of users. For each user $i=1,\dots,U$, where $U$ is the total number of users, compute the cross-correlation $\rho_{i,j}(k)$ between the trace for this user $i$ and the trace for user $j$ at lags $k=0,1,\dots,10\,\log_{10}(K/2)$, where $K$ is the number of samples in each trace.  The novelty metric $\rho_{\mathrm{nov}}$ is defined as:
\begin{equation}
  \rho_{\mathrm{nov}} = 1-\frac{1}{U}\sum_{i=1}^U \max_{\substack{j \in \{1,\dots,U\} \smallsetminus \{i\} \\ k \in \{0,\dots,10\,\log_{10}(K/2)\}}} \rho_{i,j}(k)
\end{equation}
Note that here we are not interested in matching the novelty of the source trace and the generated
trace, but in simply maximizing the value of the novelty metric $\rho_{\mathrm{nov}}$ of the set of generated traces.  A larger value of $\rho_{\mathrm{nov}}$ means that the traces generated for the different users are distinct from one another.

It is common with misconfigured and overfitted GANs to generate traces for different users that are virtually identical\footnote{This is a special case of \emph{mode collapse}, where the generator managed to produce a trace that successfully fooled the discriminator and thereafter simply repeats that single trace every time it is asked to generate one.}, and they would hence have a novelty metric close to 0 as $\max_k\rho_{i,j}(k) \approx 1$ for each $j$.  The key purpose of this metric is to enable us to avoid such a scenario and to give an indication of the uniqueness of the generated user traces. The metric $\rho_{\mathrm{nov}}$ is not part of the GAN training and hence does not need to be differentiable.  In our experiments, for batches of 100 users we computed $\rho_{\mathrm{nov}} \approx 0.7$ on the original data traces.

\subsection{Validation of generated traces}
We also note that as part of the post-validation of generated traces, we split and hold out a test data
set of users that we then compare the above three metrics to, in order to evaluate: (a) the stability across training and test data, and (b) the ability of benchmarks to predict, versus just fit, these three metrics.

Lastly, we remark that we have also tested a large number of other metrics, such as partial autocorrelation functions, Hurst exponents, and KL-divergence, but found
the above three metrics to be the most stable and indicative of desirable properties in the generated traces.

%% file: model.tex
\section{MassGAN}\label{sec:model}
In this section, we describe in detail the design of the specific GAN model proposed in the present work, which we call MassGAN.  Recall from Sec.~\ref{sec:background} that while all GAN models have the same adversarial architecture of a generator network and a discriminator network, the details of the loss functions used for training the generator and discriminator are what distinguish one GAN model from another.

\subsection{Adversarial architecture of MassGAN}
Like all GAN models, MassGAN also has a generator network and a discriminator network.  One salient feature of MassGAN is that, like the C-RNN-GAN models proposed
in ~\cite{mogren2016} on which MassGAN is based, both the generator
and the discriminator are recurrent neural networks
with long short-term memory (LSTM). Moreover, the discriminator network is a bidirectional LSTM in order to enhance its ability to detect differences between samples from the true distribution and samples simulated by the generator network.

\subsection{MassGAN discriminator loss function for training}
The discriminator loss function~\eqref{eq:gan_discloss} is unchanged from that of C-RNN-GAN~\cite[Sec.~3]{mogren2016}, which in turn is the same as for the original GAN model~\cite[Alg.~1]{goodfellow2014}.  This loss function accounts for the two objectives of training the discriminator, namely maximizing both the probability of correctly identifying samples from the true distribution and correctly rejecting samples from the generated distribution.

Just as the C-RNN-GAN modified the generator loss function from~\eqref{eq:gan_genloss} in the original GAN model, we shall also modify the generator loss function in several ways, as discussed in the next three sections.

\subsection{MassGAN generator loss function for training}
Just as in the C-RNN-GAN, the MassGAN generator's objective is to fool the discriminator into classifying the samples simulated by the generator as having come from the true distribution instead.  In the notation of Sec.~\ref{sec:background}, the generator network $G_\theta$ (with parameters $\theta$) operates on a sequence of latent vectors $z^{(i)}$ which are independent identically distributed (i.i.d.)~uniformly in $[0,1]^2$.  

\subsubsection{Discrimination loss}
The discriminator network $D_\omega$ (with parameters $\omega$) has as its output the probability that the input is classified as coming from the true distribution.  Since the input to the discriminator is the output of the generator, i.e., $G_\theta(z^{(i)})$, the generator tries to maximize the probability $D_\omega(G_\theta(z^{(i)})$, or equivalently, the generator network is trained to minimize what we call the \emph{discrimination loss}\footnote{Note that although the network being trained is the \emph{generator} network, this component of its training loss function is called the \emph{discrimination loss}, which is not to be confused with the \emph{discriminator loss} function~\eqref{eq:gan_discloss} used to train the \emph{discriminator} network.} function~\eqref{eq:gan_genloss}
\begin{equation}
  \mathcal{L}_{\mathrm{disc}} = \frac{1}{b} \sum^{b}_{i=1}\log\Big(1-D(G(z^{(i)}))\Big), 
  \label{eq:discloss}
\end{equation}
where $b$ is the batch size of sequences used during each step of
the optimization, which in our case also corresponds to the
number of users in the training data.  In other words, the loss function $\mathcal{L}_{\mathrm{disc}}$ measures the ability of the generator to induce classification errors in the discriminator. Note that for brevity we have dropped the parameter subscripts $\omega$ and $\theta$ for the discriminator function $D_\omega$ and the generator function $G_\theta$ respectively. 

Recall that the objective of MassGAN is to generate time series with temporal correlations matching those from the true distribution. Although the C-RNN-GAN works off musical score features such as pitch and volume
it is easy to map out upload and download features
to the model. In our investigation, we found that the desired temporal correlations are lost, and the novelty of the generated traces is often poor, if we simply use the discrimination loss~\eqref{eq:discloss} alone to train the generator. For example, when replaying network traffic we would expect
a high download period to be matched by a higher than usual
upload period too, but a generator trained by minimizing~\eqref{eq:discloss} alone does not have this property.

We define the new components of the overall generator loss function as follows: we first
compute the desired statistic from the original data (the true distribution).  Then we
define the generator training loss as the Euclidean distance
between the statistic computed from the true (training) data and the statistic computed from the generated trace.  This in turn is done in two parts, as discussed in the next two sections.

\subsubsection{Correlation distance loss}
Without loss of generality in the sequel, we will refer to traces with two features only, namely download and upload. It should be noted, however, that the proposed approach has been formulated for, and applies to, a trace with $N$ features (see Sec.~\ref{sec:metric_corr_dist}).

We will use the Pearson correlation coefficient $r(x,y)$ between two sequences of samples of the features $x$ and $y$ from a trace, corresponding to the download and upload samples respectively, as the metric that we want to preserve.  Following~\cite{mogren2016}, we define the sequences $x$ and $y$ themselves by their \emph{representation} in the \emph{logits} (i.e., the outputs of the last layer before the softmax layer) of the discriminator network when the input is a generated trace.  In other words, the download sample sequence $x$ for the $i$th trace in the batch (i.e., the trace of the $i$th user) is given by $x = R([G(z^{(i)})]_{\mathrm{dl}})$, where $R(\cdot)$ is the representation in the logits layer of the discriminator network, and $[G(z^{(i)})]_{\mathrm{dl}}$ is the part of the generated trace $G(z^{(i)})$ that corresponds to the download activity.  Similarly, the upload sample sequence $y$ for the $i$th trace in the batch is given by $y = R([G(z^{(i)})]_{\mathrm{ul}})$, where $[G(z^{(i)})]_{\mathrm{ul}}$ is the part of the generated trace $G(z^{(i)})$ that corresponds to the upload activity.
 
We define the \emph{correlation distance loss} function $\mathcal{L}_{\mathrm{corr-dist}}$ as the correlation-distance metric $\rho_{\mathrm{corr-dist}}$ for the single pair of features (namely the download and upload).  
From~\eqref{eq:rhocd2}, we have:
\begin{equation}
  \mathcal{L}_{\mathrm{corr-dist}} = \sqrt{\left[c_{\mathrm{target}} - \frac{1}{b} \sum^{b}_{i=1} r\bigg(R\Big([G(z^{(i)})]_{\mathrm{dl}}\Big),R\Big([G(z^{(i)})]_{\mathrm{ul}}\Big)\bigg)\right]^2}
\end{equation}
where $c_{\mathrm{target}}$ is the target Pearson correlation coefficient in the training data. In other words, the motivation behind the definition of $\mathcal{L}_{\mathrm{corr-dist}}$ is to train the generator network to minimize any mismatch between the Pearson correlation coefficient of the generated traces and the Pearson correlation coefficient of the true data.

\subsubsection{Moments distance loss}
We also want to preserve the first three moments of the true data in the generated traces -- specifically, the mean $\mu$ (i.e., the first moment), standard deviation $\sigma$ (i.e., the square root of the second central moment), and skewness $\tilde{\mu}_3$ (i.e., the third standardized moment).  To this end, we define the \emph{moments distance loss} function $\mathcal{L}_{\mathrm{mom-dist}}$ as the moments distance metric $\rho_{\mathrm{mom-dist}}$, i.e., the sum of the squared errors between each of these moments ($\mu_{\mathrm{target}}$, $\sigma_{\mathrm{target}}$, $\tilde{\mu}_{3,\mathrm{target}}$) of the true data and the estimate of this moment ($\hat{\mu}$, $\hat{\sigma}$, $\hat{\tilde{\mu}}_3$) computed from the generated traces, but using the representations of the traces from the logits layer of the discriminator network, as in the definition of the correlation distance loss function above.  From~\eqref{eq:rhomd}, we have:
\begin{eqnarray}
  \mathcal{L}_{\mathrm{mom-dist}} &=& \left[\mu_{\mathrm{target}} - \frac{1}{b} \sum^{b}_{i=1} \hat{\mu}\Big(R(G(z^{(i)}))\Big)\right]^2 \nonumber \\
  & & \mbox{} + \left[\sigma_{\mathrm{target}} - \frac{1}{b} \sum^{b}_{i=1} \hat{\sigma}\Big(R(G(z^{(i)}))\Big)\right]^2 \nonumber \\
  & & \mbox{} + \left[\tilde{\mu}_{3,\mathrm{target}} - \frac{1}{b} \sum^{b}_{i=1} \hat{\tilde{\mu}}_3\Big(R(G(z^{(i)}))\Big)\right]^2.
  \label{eqn:mdloss}
\end{eqnarray}

\subsubsection{Defining an overall generator training loss}
We observe that if either $\mathcal{L}_{\mathrm{corr-dist}}$ or $\mathcal{L}_{\mathrm{mom-dist}}$ is treated as the only loss function for training the generator network, then training converges rapidly.  However, the discrimination loss~\eqref{eq:discloss} cannot be ignored when training the generator, since we have seen that including it in the training loss function improves the novelty of the generated traces.  

Since we want to match both the moments and the Pearson correlation coefficient of the generated traces with those from the true distribution, we need to define a suitable overall training loss function for the generator network that incorporates $\mathcal{L}_{\mathrm{disc}}$, $\mathcal{L}_{\mathrm{corr-dist}}$, and $\mathcal{L}_{\mathrm{mom-dist}}$.  Our investigation of generator training loss functions defined as simple linear combinations of $\mathcal{L}_{\mathrm{disc}}$, $\mathcal{L}_{\mathrm{corr-dist}}$, and $\mathcal{L}_{\mathrm{mom-dist}}$ did not yield good matches for moments or correlations, and often got stuck at false minima.

\subsection{Generator network training method}
\label{sec:mgtrain}
Instead of searching for a definition of an overall loss function for training the generator, we propose a training method that we call \emph{conditional gradient descent}.  This method has proven effective
in yielding good matches between generated traces and training data across both correlation and moments statistics.

In conditional gradient descent, the generator loss function is defined as
\begin{equation}
  \mathcal{L}_{\mathrm{MassGAN}} = \mathcal{L}_{\mathrm{disc}} + \delta_{\mathrm{corr-dist}} \mathcal{L}_{\mathrm{corr-dist}} + \delta_{\mathrm{mom-dist}} \mathcal{L}_{\mathrm{mom-dist}}, 
  \label{eq:genloss}
\end{equation}
where we initialize $\delta_{\mathrm{corr-dist}} = \delta_{\mathrm{mom-dist}} = 1$.  In other words, we initialize the training loss function for the generator network as a linear combination of the discrimination loss, correlation distance loss, and moments distance loss.

We then follow the standard stochastic gradient descent (SGD) method of drawing random minibatches of training examples from the original traces and descending along the gradient of the sum of the three loss functions.  Note that this intuitively corresponds to descending along the gradients of all the three loss functions simultaneously.  During this initial stage of training, we continue to keep $\delta_{\mathrm{corr-dist}} = \delta_{\mathrm{mom-dist}} = 1$.  Also initialize $\mathcal{L}_{\mathrm{worst, prev}}$ to a value larger than the largest value that $\mathcal{L}_{\mathrm{corr-dist}}$ or $\mathcal{L}_{\mathrm{mom-dist}}$ can reach (see below for why this is needed).

Next, we describe when and how $\delta_{\mathrm{corr-dist}}$ and $\delta_{\mathrm{mom-dist}}$ are updated.
Either at periodic intervals, or when the total loss~\eqref{eq:genloss} is lower than 
a previous minimum, we run a 
benchmark validation on the current generator model with parameters $\theta$ as follows:
\begin{enumerate}
\item Generate a new batch of traces with the current
generator model, and also generate a batch of traces (of the same size) with the uniform-fit model described in Sec.~\ref{sec:evaluation}.
\item Compute the losses $\mathcal{L}_{\mathrm{corr-dist}}$ and $\mathcal{L}_{\mathrm{mom-dist}}$ from the two new batches of traces generated in the previous step, and update $\mathcal{L}_{\mathrm{worst, new}} = \max\{\mathcal{L}_{\mathrm{corr-dist}}, \mathcal{L}_{\mathrm{mom-dist}}\}$.
\item If $\mathcal{L}_{\mathrm{worst, new}} < \mathcal{L}_{\mathrm{worst, prev}}$ then:
\begin{itemize}
\item Save the generator model with the current parameters $\theta$ as a candidate generator model;
\item Update $\mathcal{L}_{\mathrm{worst, prev}} \leftarrow \mathcal{L}_{\mathrm{worst, new}}$;
\item For a fixed pre-set number of SGD training epochs starting from the present one, set $\delta_{\mathrm{corr-dist}}$ and $\delta_{\mathrm{mom-dist}}$ as follows:
\begin{eqnarray}
	\delta_{\mathrm{corr-dist}} &=& 1\{\mathcal{L}_{\mathrm{corr-dist}} > \mathcal{L}_{\mathrm{mom-dist}}\}, \label{eq:delta_corrdist} \\
	\delta_{\mathrm{mom-dist}} &=& 1\{\mathcal{L}_{\mathrm{mom-dist}} > \mathcal{L}_{\mathrm{corr-dist}}\}, \label{eq:delta_momdist}
\end{eqnarray}
where $1\{\cdot\}$ is the indicator function, and choose $(\delta_{\mathrm{corr-dist}}, \delta_{\mathrm{mom-dist}})$ to be either $(1,0)$ or $(0,1)$ with equal probability (by tossing a fair coin, say) when $\mathcal{L}_{\mathrm{corr-dist}} = \mathcal{L}_{\mathrm{mom-dist}}$.
\item Note that this is equivalent to dropping the statistic that is currently performing better during training (i.e., has lower training loss) from the overall training loss function~\eqref{eq:genloss} for a certain number of SGD training epochs, and descending along the gradient of the remaining statistic (the one whose $\delta$ equals unity).  Thus the two $\delta$ quantities, exactly one of which can be nonzero, act as indicator functions selecting the component loss function whose gradient is to be descended.
\end{itemize}
\item Training is complete either after a certain maximum number of training epochs, or once the condition $\mathcal{L}_{\mathrm{worst, new}} < \mathcal{L}_{\mathrm{worst, prev}}$ is no longer satisfied.
\item The candidate generator network is the trained final generator network.
\end{enumerate}

The above approach can be generalized to account for other, or additional, statistics measuring matches between generated and training traces beyond correlations and
moments. The only requirement is that the appropriate loss function for that statistic 
be efficiently differentiable so that SGD training on it is possible. 

An important aspect of the above training process is that it is done in batches
and the statistical properties of the original trace are only maintained in
subsamples of the same number of users and with the same trace lengths. In our case, we 
run with 100 batches and sequences of 12 steps. Although the same generator
can generate arbitrary-sized batches and trace steps, in all the numerical results reported in the present work, the statistics are always validated using these same batch and sequence values.

\subsection{Transfer Learning for Context-Aware GAN}
We want to generate traces that are \emph{context-aware}, where for the specific application of user activity trace generation, the \emph{context} is a categorical parameter describing an attribute or property of all or part of the user activity recorded in the trace, and which cannot be modified by the network or the trace measurement procedure.  Although users' activity (as recorded in traces) may vary greatly depending on a variety of factors, without loss of generality in the present work we focus on two of these factors, namely the signal strength and the type of application, as they significantly affect user data usage.  

In other words, in the present work we define the context to be either the signal strength (\texttt{HIGH} or \texttt{LOW}) or the type of application being run (\texttt{STREAM} or \texttt{INTERACT}), or a combination of signal strength and application type (\texttt{STREAM\_HIGH}, \texttt{STREAM\_LOW}, \texttt{INTERACT\_HIGH}, or \texttt{INTERACT\_LOW}).

The training dataset we use, namely the Telefonica Mobile Phone Use Dataset (see Sec.~\ref{sec:data}) has no contextual labels, which requires us to infer context from the data itself.  To create a context-aware GAN we split the training data traces into contextual cohorts corresponding to the 8 context labels defined above (\texttt{HIGH}, \texttt{LOW}, \texttt{STREAM}, \texttt{INTERACT}, \texttt{STREAM\_HIGH}, \texttt{STREAM\_LOW}, \texttt{INTERACT\_HIGH}, and \texttt{INTERACT\_LOW}), where we assign a context label to all or part of a trace based on received signal strength and application type.  We also check that each collection of traces has enough entries in terms of users and sequence length of traces\footnote{In the present work, in order for us to train a model there need
to be at least 5 users with at least the configured sequence length number of
data points (12 in our case).}.  More precisely, we cluster the split traces into 8 clusters, one for each context label, based on \emph{dissimilarity metrics} which we define to be the relative difference in means between the collections of traces assigned to the different clusters, and the relative difference between the mean of the traces in a given cluster and the mean of the full training data.  We expect this context labeling to yield collections of traces with qualitatively different characteristics. 

Next, we use techniques of transfer learning~\cite{tan2018} to train a GAN model for each of the 8 contexts defined above, by starting with the ``global'' GAN model trained on the entire training data using the algorithm described in Sec.~\ref{sec:mgtrain} and then ``fine-tuning'' that model by training on the collection of traces with that context label for a certain number of training epochs (but a fewer number of training epochs than employed when training the global model).  In other words, we now have not one generator network (corresponding to the generator of the global GAN model) but a whole family of generator networks, comprising the global GAN model's generator and one generator for each context defined above. 

Note that our approach above created a total of 8 contexts from two context categories (signal strength and application type) with two groups in each category (\texttt{HIGH}/\texttt{LOW} for signal strength, \texttt{STREAM}/\texttt{INTERACT} for application type).  Although this approach generalizes to quantizing into more than two groups in two contexts, it quickly becomes inefficient if we have too many contexts stemming from too many combinations of groups and/or context categories.

In the latter case, more aggressive filtering of the traces in the ensuing collections may have to be performed, possibly involving not just the mean $\hat{\mu}$, but also the standard deviation $\hat{\sigma}$ and/or the skewness $\hat{\tilde{\mu}}_3$.  Alternatively, the threshold for difference between the means computed over the traces assigned each context label may be set to be higher than the value of 10\% that we set in the present work\footnote{The threshold may be treated as a \emph{hyperparameter} that is optimized through an outer iterative loop as part of an AutoML workflow~\cite{hutter2018}, but we do not elaborate on this here.}.
More discriminative splits of the training data across the different context labels will also lead to many context label trace collections failing the threshold of the minimum users with the minimum length sequences assigned that context label.

If a sufficient (in size and quality) training dataset does not exist for a particular context, then we cannot fine-tune the global GAN model for this particular context by training it on traces that are assigned this context.  In that case, when the family of generators is deployed for context-aware trace generation purposes and a context is specified for which the corresponding specialized generator has not been trained, then we just run the global generator.


%% file: data.tex
\section{Network Traffic Data}\label{sec:data}
Our approach of realistic user traffic trace
replay is rooted in mimicking (without copying exactly) original
traces measured with real user traffic.

Any trace that has a timestamped series of
upload-, download volumes and signal strength
values grouped by users could be applied.

Here, we use a trace collected by Telefonica
in Spain in 2016, called the Mobile Phone Use Dataset~\cite{pielot2017}. 
The dataset contains traces from more than 300 mobile phone users over
several weeks. Every 10 minutes a sample is taken where the transmitted
and received data over the last 10 seconds are recorded with the current
Wi-Fi and LTE signal strengths. We notice that the Wi-Fi data transmissions
are more prevalent and thus focused on those in order to capture interesting
behavior between signal strength and data volume. The traces also capture
the application that is at the top of the stack, i.e. currently or latest
used.

We use Wi-Fi signal strength to partition data where all time steps with
RSSI $<-75$ get labeled with the signal strength \texttt{LOW} context and the
rest are labeled \texttt{HIGH}. 

For applications we first map each application into its official app store
category, and then we map all applications that fall into one of the categories:
{\it MUSIC\_AND\_AUDIO}, {\it MAPS\_AND\_NAVI-GATION}, {\it SPORTS}, or {\it VIDEO\_PLAYERS}
into the \texttt{STREAM} context. All other apps are mapped to the \texttt{INTERACT}
context.

Given that the 10\,s sample every 10 minutes is noisy we average 6 values to
produce hourly samples. For applications we create a memory where the last
known application that could be successfully mapped to a category like above
becomes the app for that hour.

We look at sequences of 12h, and any users whose traces are less than 10h after
a context split are discarded from the 50/50 user split training and test data sets.
For the global context we gather 100 users in each data set, a context split that
has less than 5 users is also discarded.

The context splits are as follows. Hourly values with application \texttt{STREAM} vs \texttt{INTERACT},
and signal \texttt{LOW} vs \texttt{HIGH}, and any combination of the two contexts are checked whether
they pass this cutoff. If they do we compare the difference in mean download and upload rates
between the global context (all values) and the context split. If it is more than $10\%$ we call
that context significant and keep it, otherwise the context is discarded.

The resulting significant contexts are trained separately and obtain separate GAN generator
models. The insignificant contexts that were discarded can still be used when requesting a
context trace through an API that we will discuss later, but they are generated from the
default global context.

For comparison and illustration of generality of the model training process we
also use a second data set collected from 200 cable modems in a residential
area over the period of a month. The advantage of this data
is that it is very fine grained and time synchronized. We collect 6-minute
upload and download volumes averaged over minute-by-minute samples on each
modem separately~\footnote{This is the same level of aggregation (6-values) as with the Mobile Phone trace.}. 
More information about the data can be found in~\cite{sandholm2019}.  
However, since we do not have app or signal information in the data, we can only use the data to train a context-unaware model. 

\begin{figure}[htbp]
\centering
        \includegraphics[scale=0.5]{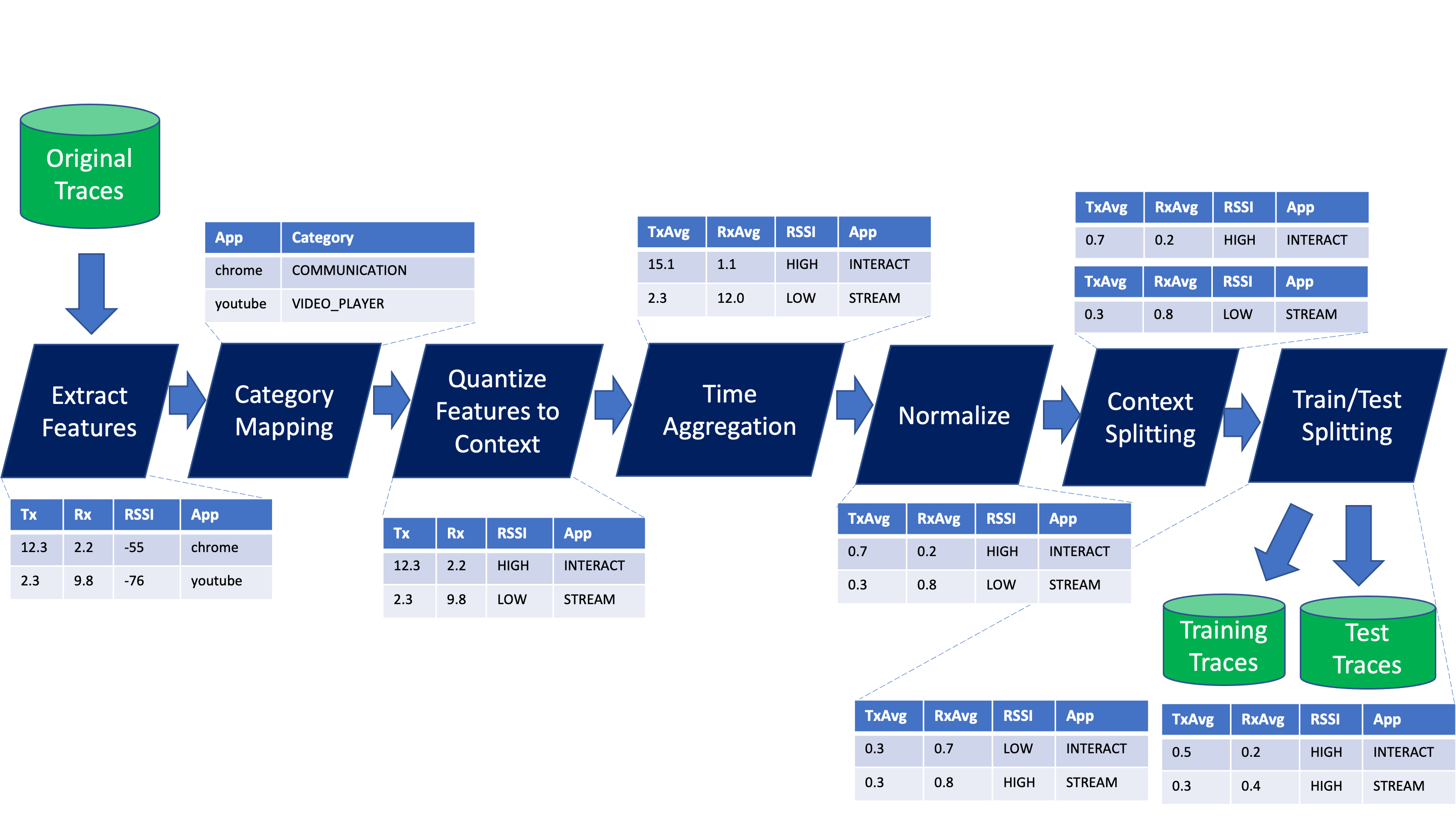}
	\caption{Pre-Training Data Transformation Pipeline.}
	\label{datapipe}	
\end{figure}

The pre-training data transformation pipeline is depicted in Figure~\ref{datapipe}.

%% file: evaluation.tex
\section{Model Evaluation}\label{sec:evaluation}
We use the three previously mentioned metrics, 
{\it correlation}, {\it moments} and
{\it novelty} to evaluate our GAN model against three
benchmarks, {\it Uniform Fit (Uni) }, {\it Generic Distribution Fit (Dist)} and
{\it timeseries GAN (TimeGAN)}, discussed next. 

{\bf Uni}. The uniform fit model simply computes the min and max of
a time series to reproduce and then generates a new time series
with a uniform distribution within that range.

{\bf Dist}. The generic distribution fit model evaluates over 100
distributions of different characteristics and picks the best fit
with the Generalized Additive Models for
Location, Scale and Shape (GAMLSS) proposed in ~\cite{rigby2005} and
implemented in the gamlss R package~\footnote{https://www.gamlss.com/}.

{\bf TimeGAN}. The TimeGAN benchmark uses the method proposed in~\cite{yoon2019}.
To use the model we map each user trace of download and upload samples to
a couple of new features, and ensure all user traces have the same sequence length.
Note this only allows us to generate the same number of users as in the input,
which for the following evaluation is done for all benchmarks.

Here we only compare the global, context-unaware models.
We split the data in 50/50 between train and test users with
100 users in each group. In the evaluation we compare the
generated traces both to the original training data and
the test data that was hidden during training.

\begin{table}[htbp]
        \caption{Benchmark Metrics Mobile User Trace Evaluation Results.}
\begin{center}
\begin{tabular}{|l|l|l|l|l|}
\hline
\textbf{Benchmark} & \textbf{Data} &\textbf{Correlation} & \textbf{Moments} & \textbf{Novelty} \\
 &  &\textbf{Distance} & \textbf{Distance} &  \\
\hline
\textbf{Uni} & train & .48 & 6.0 & .55 \\
            & test & .49 & 5.6 & \\
\hline
\textbf{Dist} & train & .42 & 3.9 & .28 \\
             & test & .44 & 3.5 & \\
\hline
\textbf{TimeGAN} & train & .08 & 4.8 & .39\\
                 & test & .10 & 4.3 & \\
\hline
\textbf{MASS} & train & .04 & .21 & .42 \\
              & test & .06 & .62 & \\
\hline
\end{tabular}
\label{T:benchmark}
\end{center}
\end{table}

\begin{table}[htbp]
        \caption{Benchmark Metrics Cable Modem Trace Evaluation Results.}
\begin{center}
\begin{tabular}{|l|l|l|l|l|}
\hline
\textbf{Benchmark} & \textbf{Data} &\textbf{Correlation} & \textbf{Moments} & \textbf{Novelty} \\
 &  &\textbf{Distance} & \textbf{Distance} &  \\
\hline
\textbf{Uni} & train & .24 & 8.1 & .56 \\
            & test & .22 & 8.3  &  \\
\hline
\textbf{Dist} & train & .22 & 5.3 & .16 \\
             & test & .21 & 5.4 & \\
\hline
\textbf{TimeGAN} & train & .01 & 7.5 & .52 \\
                 & test & .03 & 7.7 & \\
\hline
\textbf{MASS} & train & .003 & .63 & .53 \\
              & test &  .01 & .79 & \\
\hline
\end{tabular}
\label{T:benchmarkcable}
\end{center}
\end{table}

We can see from the results in Tables~\ref{T:benchmark} and~\ref{T:benchmarkcable} 
that MASS 
dominates the other benchmarks in terms of {\it correlation} and
{\it moment} fits across both the original data and the test data
prediction. TimeGAN also reproduces correlations well, but has the limitation
of fixing the number of users that can be generated to the original
set, and does not reproduce moments well. Interestingly we also see
that novelty suffers if we try to make the best possible distribution
fit. Recall that a high novelty score allows us to expand to more
generated users beyond the original data set without ending up re-generating
the same traces. Novelty is defined in terms of cross-correlations
within the generated trace so it will be computed the same way
regardless of whether you compare it to the training user data or the test 
users, which is why it only has a single value per benchmark in the table.

The lower correlation distances in the cable modem data set can be explained
by there in general being a lower correlation between download and upload
traffic volume compared to in the mobile user data set\footnote{Although lower, the correlations are still significant and around $.25$ compared to about $.5$ for the Moble Phone data.}. Another difference
between the data sets is that the moment fit is worse across the board for
the cable modem trace, which could indicate that it is more noisy data,
due to the shorter time frame it was aggregated over (6 minutes versus an hour).

We note here that there are analytical models for distribution fits
with preserved correlations between time series known for Gaussian
and generalized beta distributions~\cite{olkin2015,lin2016}, 
but not for general distribution fitting.
The general distribution fitting did surprisingly
poorly in reproducing moments compared to MASS. 

All these test were run
with 2000 epochs. We will evaluate the impact the epochs have on the
results later on.

\subsection{Context Evaluation}
Next, we compare the metrics performance of contextual
models versus global models to determine
how valuable context-based generation is.
We compare the original contextual split
to the contextual generator as well as
a non-contextual generator and compute
$\frac{m-m_c}{m}$ as the proportional
change,
where $m$ is the global value and $m_c$
the context aware metric. For the
distance metrics (correlation, moments)
a positive value means a reduction in
distance, which is expected and good. However,
a higher novelty value is good, so a negative
value for novelty in this comparison implies
an improvement. 
The novelty metrics is not expected to improve
as a specialization into context could potentially
reduce the novelty of the generated trace, in particular
if the number of users matching the context is low.

The results are shown in Table~\ref{T:context}.
Note that we only run these evaluations for
the contexts that were deemed significant, as
they are the only ones with a trained generator. 

\begin{table}[htbp]
        \caption{Context Proportional Change.}
\begin{center}
\begin{tabular}{|l|l|l|l|}
\hline
\textbf{Context} & \textbf{Correlation} &\textbf{Moments} & \textbf{Novelty} \\
 &\textbf{Distance} & \textbf{Distance} &  \\
\hline
\textbf{LOW} & .43 & .82 & -.31 \\
\hline
\textbf{INTERACT\_LOW} &  .33 & .81 & .52 \\
\hline
\textbf{STREAM} & .47 & .97 & -.09\\
\hline
\textbf{STREAM\_HIGH} & .42 & .96 & .05 \\
\hline
\textbf{STREAM\_LOW} & .82 & .82 & .85 \\
\hline
\end{tabular}
\label{T:context}
\end{center}
\end{table}

It is clear that there is a tradeoff between
better moment and correlation matching versus
reduced novelty in some cases. Therefore, based on this analysis 
further contexts may be pruned, like \texttt{STREAM\_LOW},
and \texttt{INTERACT\_LOW}, which have very high degradations in novelty.

\subsection{Learning Evaluation}
\begin{figure}
	\centering
	\begin{subfigure}[b]{0.49\textwidth}
		\includegraphics[width=\textwidth]{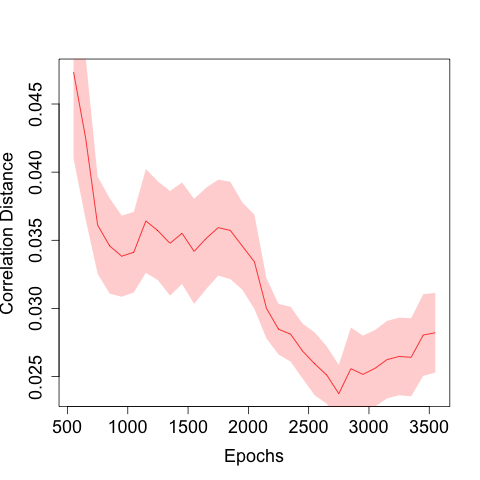}
	\end{subfigure}
	\begin{subfigure}[b]{0.49\textwidth}
		\includegraphics[width=\textwidth]{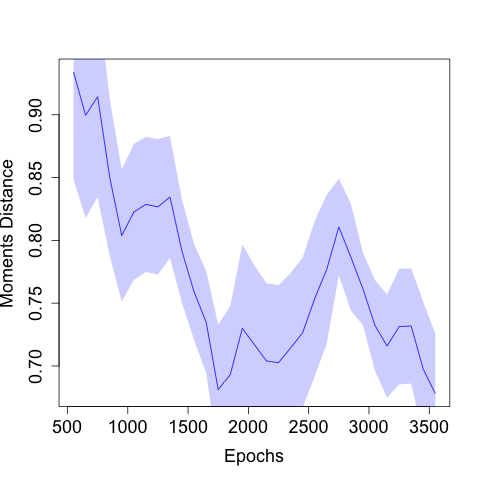}
	\end{subfigure}
	\begin{subfigure}[b]{0.49\textwidth}
		\includegraphics[width=\textwidth]{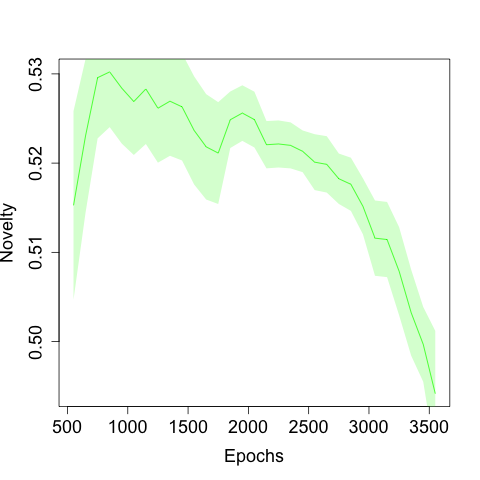}
	\end{subfigure}
	\begin{subfigure}[b]{0.49\textwidth}
		\includegraphics[width=\textwidth]{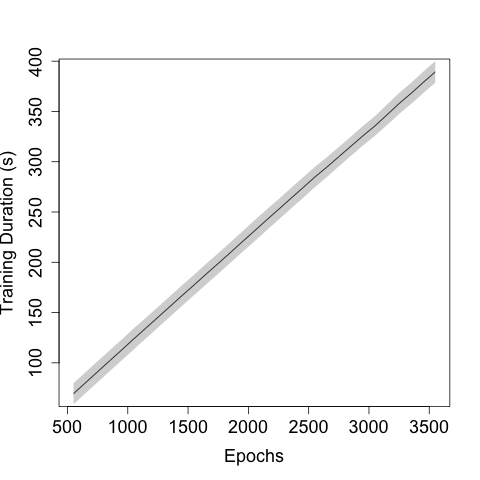}
	\end{subfigure}
	\caption{Learning Curves for Metrics and Training Duration}
	\label{learningcurve}
\end{figure}

Next, we study the time it takes to train the different
metrics. Note here that we only have loss functions
that try to minimize correlation distance and moments
distance and that novelty is only indirectly kept
in check. From Figure~\ref{learningcurve} we can
see that the moments distance reaches its minimum
after about 1800 epochs, whereas the correlation distance
reaches its minimum after about 2800 epochs. In training time
that is about 3 minutes versus 5 minutes~\footnote{The training
was run using an NVIDIA GeForce RTX 3060 6GB RAM, 
1.785GHz GPU with CUDA on an 8 quadcore CPU 1-1.4Ghz
Linux Ubuntu 20 desktop PC.}. Note that these two metrics
are jointly optimized with the GAN discriminator loss. Training
each metric separately converges much faster, but is less
interesting for our purpose.
Novelty as expected has very minimal variation regardless of
training time. There seems to be a marginal downwards trend
in novelty with increased training time, but the range
of novelty values produced (y-axis) is very small ($.515-.530$). 
A downwards trend in the epoch range of $1500-2800$ in correlation
distance seems to coincide with an upwards trend in moments distance,
highlighting that minimizing for both of these metrics at the
same time is a challenge and results in various trade-offs having
to be made, in terms of which of the metrics is the most important
to get under a given level. Currently we optimize both without any
bias or weights. Another observation is that there seems to be
a plateau around epochs $1000-2000$ for the correlation metric,
which we eventually overcome. These kind of plateaus is the reason
we employ our conditional gradient descent mechanism. 

The graphs are using rolling means
of 10 periods to smoothen the curves, and make trends more
apparent. We also ran each training epoch length in five repeats
and then took the average value for all metrics. The standard
error regions are calculated from the same 10-period windows as
the means.

%% file: system.tex
\section{System Design}\label{sec:system}
Next, we discuss how training and replay was implemented
and integrated into various networking tools.

\begin{figure}[htbp]
\centering
        \includegraphics[scale=0.5]{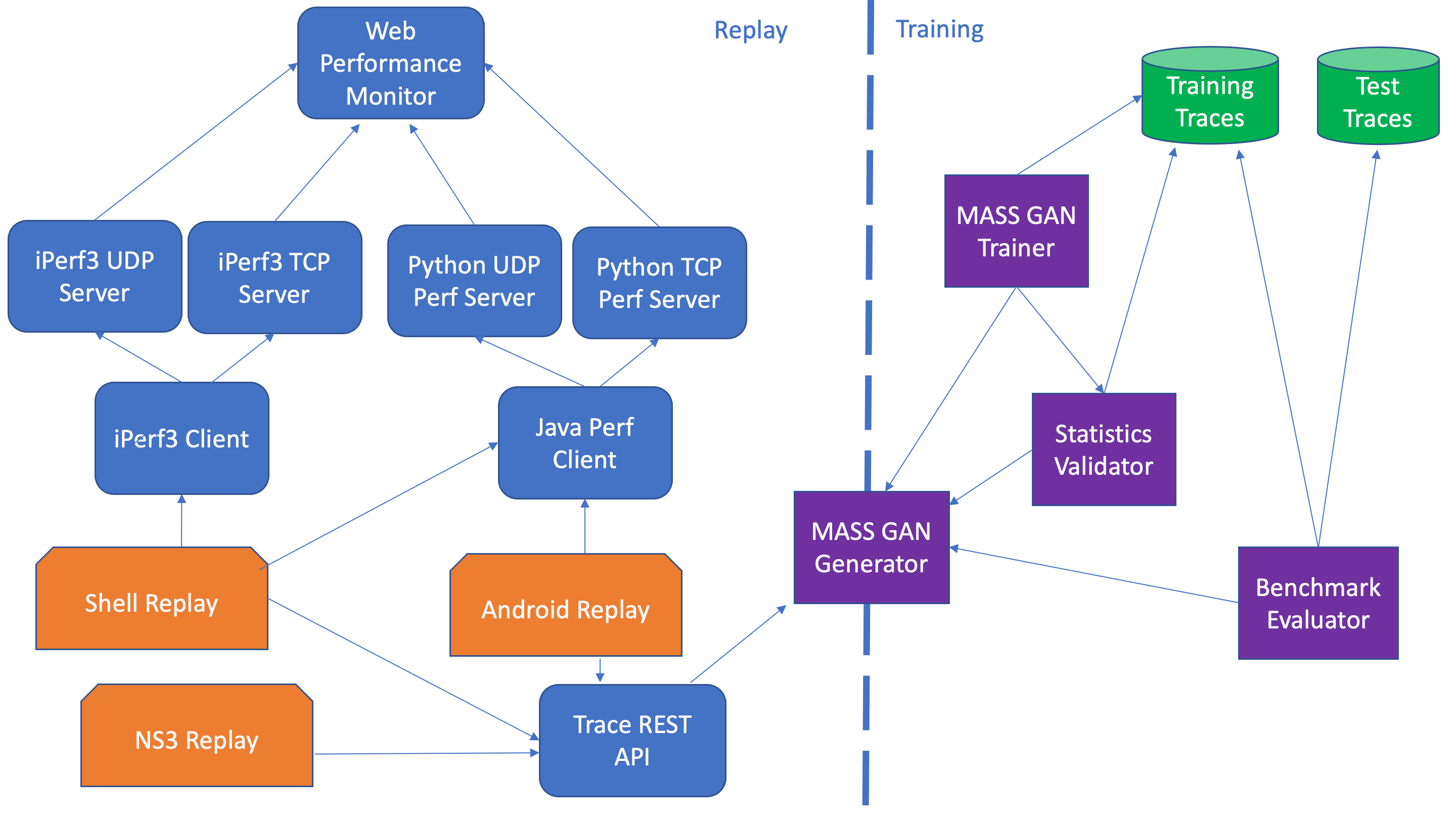}
	\caption{Training and Replay Component Interactions.}
	\label{massinteract}	
\end{figure}

The training and replay component interactions
are shown in Figure~\ref{massinteract}.

\subsection{GAN Training}
The GAN implementation is built on top
of a PyTorch port of C-RNN-GAN~\footnote{https://github.com/cjbayron/c-rnn-gan.pytorch}.
The discriminator model is used as is, but the generator was modified
with new loss functions. The training process was rewritten as well as all data loading and
validation functions.

The validation process as well as the basic distributional benchmarks are implemented 
in R. The TimeGAN benchmark was used without code modifications, again using
a PyTorch port~\footnote{https://github.com/d9n13lt4n/timegan-pytorch}.

Training is run on a GPU via CUDA drivers to speed up the process.

Although Python and PyTorch can be installed on most platforms without
GPU support, we also wanted to target embedded platforms with limited
resources. Even if the tools are available, making sure the right versions
are installed and being abe to import the saved models on many different
platforms is a cumbersome undertaking. Hence, we implement a simple REST API
with both JSON and clear text output to generate traces on demand from a
pre-trained set of models.

Although the models take a long time to train on a CPU as opposed to a GPU
they may easily be used
for execution anywhere. The saved model files ($\sim4\,\text{MB}$ per generator) can be transferred
to the web server hosting the REST API if it is a different machine than where
training took place. 

\subsection{Shell Replay}
To replay traces we, optionally as a drop-in replacement for iPerf3, 
implement a custom UDP and TCP socket client-server tool
that behaves similar to iPerf3 in that it allows for transmission of randomly generated
data at given rates with specified message sizes on both upload and download directions
across a pair of end-points. The performance servers also log their performance history,
which in turn is picked up by the REST server and served using html5 canvas charts for easy 
real-time monitoring. The performance servers may be deployed anywhere in your network
and do not have to be collocated with the REST server generating traces.

The reason for not mandating use of iPerf3 is to 
simplify use from platforms where iPerf clients
are difficult to use, e.g. Android. The client is written in Java without
any dependencies beyond the JRE and can run both inside an Android app and 
standalone where a JRE is available. The server is written in Python. A simple protocol
was designed to let clients communicate: 1) direction, 2) duration, 3) rate
and 4) message size of the replay. Separate endpoints are also given for UDP
vs TCP. Endpoints like iPerf3 can only serve a single stream at a time, so 
separate endpoints are used to replay upload and download traffic concurrently.

On some constrained devices, e.g. OpenWRT routers, it may however be easier to install
iPerf3 than a Java runtime. So we also implement a hook to 
replay traffic through iPerf3 if available. Apart from iPerf or a JRE to
replay traces and curl to generate traces via the REST API, the
script has no other dependencies other than Bash, so it should be easy to run
from a multitude of platforms.

The replay script can be controlled with environment variables that are sourced
from a file before a replay starts. Variables that can be configured include
MASS server and performance server endpoints, maximum upload and download 
rates~\footnote{The trace generator produces normalized rates that need to be multiplied with
the desired maximum rates.}, duration of each replay time step called epoch, and app context
transition probabilities. The current Wi-Fi signal strength may be used to drive the context
trace selection as well. If the currently connected Wi-Fi network has an RSSI less than
-75 a trace with the \texttt{LOW} signal context will be replayed. The probability of
a replay step being replayed with UDP as opposed to TCP traffic may also be set.

A replay run is defined by how many steps it replays, and the same number of steps are
requested from the trace generation API for each known context that is supported by
the client. The sequence could be replayed multiple times or whenever a sequence has
finished replaying a new trace sequence may be downloaded. During replay the current
app and signal contexts are obtained and the corresponding context trace is then picked
to replay the upload and download rates of the current epoch. This pre-caching of
context traces allows for quick jumping between parallel traces without having to
interact with the trace REST API for each step, and thus avoiding any impact to 
the performance of the replay.

\subsection{Android Replay}
We developed an Android app that allows for configuration and replay similar to the
shell replay script described above.
The signal context is inferred from the current cellular provider or the currently
connected Wi-Fi network. The Java rate replay client is embedded in the app through
a native Java API and can run both in the emulator during development and on real devices.
The idea here is to allow a single app to mimic the network behavior of many apps
while still reacting to changes in RF bandwidth e.g. due to mobility or obstruction.
This design avoids having to script or automate launching of real apps during
an experiment to produce realistic network traffic.
The backend servers of an Android replay are identical to the shell replay so
the same monitoring tools may be used, and real UDP and TCP traffic will traverse
the Wi-Fi or cellular network.
The app may be downloaded from an app store or be side-loaded, and does not require
any custom deployment or root access. In terms of dependencies it only relies on the
Volley and JSONObject libraries for the REST interaction with the trace generation API.

\subsection{NS3 Simulated Replay}
Finally, we also implemented an integration with NS3 to allow you to replay MASS
traces inside your simulated NS3 networks. We have tested the integration
with Wi-Fi and LTE network simulations, but it should work with
any network where you have at least a pair of endpoints reachable by IPv4
addresses.

We have developed a custom UDP and TCP client NS3 Application that sends variable rate
data based on MASS traces and contexts such as signal strength and app type (\texttt{STREAM} or \texttt{INTERACT}).
Transmissions are done epoch by epoch with upload and download replays in
parallel, just like in the real replay clients described above.

Traces are pre-generated for each supported context before a simulation is started
and then pulled in based on the current context. We developed a C++ API to install
a set of apps on a pair of endpoints from a Node container and IPv4 interface container
inside the NS3 simulation code.
Each endpoint deployed may be configured separately with parameters identical to
the Android and shell clients, such as UDP or TCP traffic, max download and upload
rates etc.

We use the trace source callback mechanism to detect changes in signal strength from Wi-Fi and LTE
PHY stacks that are detected automatically. The sources we listen to include MonitorSnifferRx
and ReportUeMeasurements for Wi-Fi and LTE PHYs respectively. RSSI less than $-75$ or RSRQ
less than $-15$ are mapped to the \texttt{LOW} signal strength context.

%% file: usecase.tex
\section{Use Case: Wi-Fi Contention Window Control}\label{sec:usecase}
To illustrate how to use MASS, we now
describe an experiment that evaluates an ML-based Wi-Fi Contention Window 
Backoff algorithm, called MLBA~\cite{sandholm2019}, designed to set contention 
window ranges according to traffic load.

In this particular example we want to investigate the fairness between
upload streams and download streams under contention across interfering
APs in a Wi-Fi 5 versus a Wi-Fi 6 setup, with our custom contention
window algorithm versus the default backoff algorithm on the
chipset under investigation, called here simply BA.

The context-aware MASS GANs for the Mobile Phone Use data set were used in this
case and deployed behind a Web server in the Wi-Fi test bed. 

Trace replay is based on the live sensed signal (RSSI) and for simplicity 
we only make use of the \texttt{STREAM} application context in this 
example~\footnote{If we expect different traffic behaviors like interactive
and batch apps to impact the results we could add that as a condition too.}.

We have 4 STAs each connected to a dedicated AP. The APs run iPerf servers to allow
traffic generation. All APs are set up on the same channel and both STAs and APs are
positioned in close proximity to each other to ensure there is interference.

The MASS shell script is deployed on all STAs along with an iPerf client.

Each STA generates
a trace independently, and traces are reused across experiment condition, i.e. with and
without our controller. 

For each iteration we first replay the traces against the default backoff algorithm,
and measure upload and download throughput achieved and compare them to the intended
throughputs from the trace to measure bias.

Next we replay the traces during a calibration phase where we train our ML algorithm,
and then finally we run our ML algorithm in prediction or execution mode, measuring
the throughput and bias again.

Both APs and STAs have identical hardware and software stacks (OpenWrt 21.02), 
and there is no bias in positioning between APs and STAs in the testbed.

We use the HE80 (Wi-Fi 6 80MHz band) and VHT80 (Wi-Fi 5 80MHz band)
modes and set both max upload and download rates to 300Mbps.
The single STA to AP and AP to STA stream throughput without 
interference was measured to be between 300 and 400~Mbps.

The range of contention windows (CWMIN,CWMAX) we train with is between 
sizes 7 and 127.

We measure bias and aggregate throughput in epochs of 5, with 30
unique traces generated by each STA (total of 120 users simulated).

Bias is defined here as:
\begin{equation}
\frac{\hat{tp}_{up}}{\hat{tp}_{down}} - \frac{{tp}_{up}}{{tp}_{down}}  
\end{equation}

where $\hat{{tp}}$ is the throughput measured in the given direction
that is attempted based on the trace generated and ${tp}$ the measured
throughputs.

Given this definition we observe that a bias that is positive means
download traffic is getting a higher share of bandwidth than intended,
we call this a download bias. Conversely, if the bias is negative it means
that upload traffic is prioritized and we call this an upload bias.
Clearly the closer the bias is to zero the more fair the bandwidth allocation
is.

\begin{figure}
	\centering
	\begin{subfigure}[b]{0.49\textwidth}
		\includegraphics[width=\textwidth]{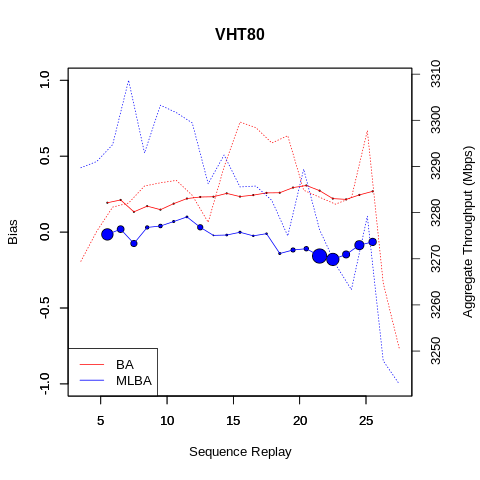}
	\end{subfigure}
	\begin{subfigure}[b]{0.49\textwidth}
		\includegraphics[width=\textwidth]{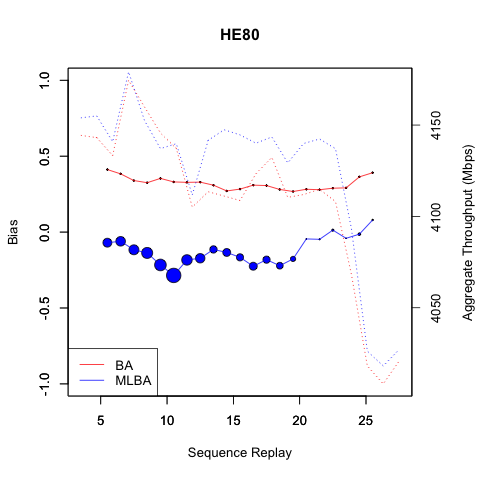}
	\end{subfigure}
	\caption{Bias (solid lines, left side y-axis) and Throughput (dotted lines, right-side y-axis)}
	\label{bias}
\end{figure}

Figure~\ref{bias} shows the results for VHT80 and HE80.
Throughput values are aggregate Mbps achieved across all STAs within 
a sequence replayed. Each sequence is 5 epochs. The graphs show
rolling averages of 10 sequence replays. The dotted lines show throughput
on the y-axis on the right. The filled circles on the MLBA line are
proportional in size to the contention window selected for the sequence. 
In general, a lower CW is selected with a lower load. This is most
prominently on display towards the end of the 30 sequences run in the HE80 experiment, 
where the load goes down, and a lower CW is also selected (smaller circles)\footnote{We note here
that there is no expectation that the learning will improve over the 30 iterations as each
iteration has its own independent training and prediction phases.}.

Looking at pairwise t-tests we
observe that bias was reduced and the reduction is statistical significant
on a 95\% significance level. The difference in throughput, was however
not significant. The means are shown in Table~\ref{T:bias}.

We also observed that the download bias is stronger in HE80 compared to
VHT80 with the default backoff algorithm ($~20$\% vs $~40$\%), and to compensate for that
we give our ML algorithm a lower cutoff in contention range probed (50
instead of 127). The throughput, as expected is also about 25\% higher
with HE compared to VHT.

\begin{table}[htbp]
        \caption{Mean Bias and Throughput Results Summary.}
\begin{center}
\begin{tabular}{|l|l|l|l|l|}
\hline
\textbf{Benchmark} & \textbf{Protocol} &\textbf{Bias} & \textbf{Throughput} \\
\hline
\textbf{BA} & VHT80 & .23 & 3273  \\
            & HE80 & .36 & 4091\\
\hline
\textbf{MLBA} & VHT80 & -.027 & 3273 \\
             & HE80 & -.052 & 4108 \\
\hline
\end{tabular}
\label{T:bias}
\end{center}
\end{table}

One could argue that changing the EDCF transmission queue parameters
on the APs also could have achieved increased fairness, but clients may come and
go and emit variable load, making it non-trivial to determine what is optimal. 
The point of these type of ML algorithms in general is to avoid
having to change configuration settings manually when load conditions change.

Learning the load and evaluating a learning algorithm is a challenge without
realistic user traces. Here, MASS allowed us to investigate the difference
between the demand (original trace values) and achieved value (throughput/goodput).

%% file: discussion.tex
\section{Discussion and Conclusion}\label{sec:discussion}
Traditional user load simulation techniques based on simple
statistical models are increasingly inadequate for
evaluating and testing complex models to optimize network operations.  This is especially true for machine learning models like deep neural networks.

An inability to introduce a machine learning model to sufficiently realistic scenarios during the training phase may lead to a false sense of confidence in the model's performance on the part of researchers.  This may result in disappointment and a feeling of wasted time and effort when the trained model shows surprisingly poor results in a production deployment.

GANs using RNNs have been proposed in several fields (starting with medical applications, see~\cite{esteban2017}) as a means to generate realistic time series training data for machine learning model training for applications where real-world data is scarce or expensive to collect.  The machine learning workflow is called TSTR, for ``train on synthetic, test on real'' (data)~\cite{esteban2017}.

The present work proposes a specific GAN model, which we call MASS, to address the machine learning model training needs of network operators.  We believe the ability to validate an increasingly
automated, machine-learning driven network infrastructure
calls for more sophisticated ways of simulating real user
traffic, and that GANs in general and MASS in particular
provide a good starting point.

MASS was designed with user privacy in mind, as opposed to
being an anonymization afterthought or by obscuring true behavior
with noise, and hence it is ideally suited for sharing models trained
on private data in various wireless networking scenarios,
such as residential MDU Wi-Fi or Enterprise 5G or CBRS networks
to name a few. In addition it was designed to be cognizant
of the environment the trace is deployed in by autonomously
reacting to environmental signals such as RF quality and the task
at hand, for example the application running on the UE or mobile
station. 

We envision the use of MASS in situations where some party
who traditionally has not been willing to share raw data
because of privacy concerns may be more willing to
share GAN models of the data. The level of privacy
protection for the original users then depends on
the number of users producing the original data employed
to train the GANs. It is clear that only having
a handful of users train the GANs may lead to GANs
that mimic the individual behavior of these few users
to the point where the generated traces may reveal
more private behavior than desired. It is thus
important from a privacy perspective that data be
collected from a sufficiently
large number of users to train the GANs. 

We also note
that our GAN models do not produce absolute time
series of user load, but normalized load. Each
user's load is normalized for both uploads and downloads
in the range $\{0,1\}$, and the GAN generated trace
would also then produce a value in the same range.
When an experimenter uses a GAN, max upload and download
rates have to be specified to scale the replay to the
testbed used. This normalization helps with training the
GAN more efficiently and makes it
easy to scale the traces to the target system,
but it also provides another level
of privacy protection.

In conclusion, we have seen that our trace generation method
is able to produce novel, realistic network traffic
traces that can easily be integrated in simulations as well
as experiments in testbeds.

%% file: acknowledgments.tex
\section*{Acknowledgments}
We would like to thank Irene Macaluso, Lili Hervieu,
Bernardo Huberman, and Aaron Quinto for their
feedback and reviews of different sections
of this paper.

%% file: bashoptions.tex
\section{Shell Replay Options}\label{bashoptions}
\lstset{language=bash,basicstyle=\scriptsize\ttfamily,showspaces=false,showstringspaces=false} 
\begin{lstlisting}[commentstyle=\color{blue}\textit, backgroundcolor = \color{lightgray}, framexleftmargin = 0.2em, framexrightmargin = 1em, multicols=2]
# Host with MASS REST API that should
# be used to generate traces.
MASS_HOST=localhost
# Host where perf servers are running.
PERF_HOST=localhost
# The length of generated sequences. 
# If continuous replay is not enabled 
# this is also the length of the replay.
SEQ_LEN=10
# Max download rate in Mbps.
MAX_DOWN=1
# Max upload rate in Mbps.
MAX_UP=1
# Message buffer size in replay.
BUFFER=1024
# Port where download perf server runs.
DOWN_PORT=5557
# Port where upload perf server runs.
UP_PORT=6666
# Time to replay each entry in the trace. 
# in seconds.
EPOCH_TIME=5

# Type of app to simulate at the beginning.
INITIAL_CONTEXT="INTERACT"
# Probability to stay with an interact type app 
# in the next time step if the current time step 
# simulates an interact app.
INTERACT_STAY_PROB=0.5
# Probability to stay with a stream type app in
# the next time step if the current time step 
# simulates a stream app.
STREAM_STAY_PROB=0.5
# If enabled (1) use wifi signal streanght as
# context for trace generation.
# Disable with 0.
USE_SIGNAL=1
# Probability of generating a trace entry with
# UDP.  If not then use TCP.
UDP_PROB=0.5
# After replaying a full sequnce regenerate a 
# new sequence.
CONTINUOUS=0
# Use iperf clients and servers instead of the 
# custom servers.
USE_IPERF=0
\end{lstlisting}

%% file: restapi.tex
\section{REST API}\label{restapi}

\subsection*{URL}
\lstset{basicstyle=\scriptsize\ttfamily,showspaces=false,showstringspaces=false} 
\begin{lstlisting}[commentstyle=\color{blue}\textit, backgroundcolor = \color{lightgray}, framexleftmargin = 0.2em, framexrightmargin = 1em]
/generate
\end{lstlisting}

\subsection*{Method}
\lstset{basicstyle=\scriptsize\ttfamily,showspaces=false,showstringspaces=false} 
\begin{lstlisting}[commentstyle=\color{blue}\textit, backgroundcolor = \color{lightgray}, framexleftmargin = 0.2em, framexrightmargin = 1em]
POST
\end{lstlisting}

\subsection*{URL Params}
\lstset{basicstyle=\scriptsize\ttfamily,showspaces=false,showstringspaces=false} 
\begin{lstlisting}[commentstyle=\color{blue}\textit, backgroundcolor = \color{lightgray}, framexleftmargin = 0.2em, framexrightmargin = 1em]
format=[text|json]
\end{lstlisting}

\subsection*{Payload}
\lstset{basicstyle=\scriptsize\ttfamily,showspaces=false,showstringspaces=false} 
\begin{lstlisting}[commentstyle=\color{blue}\textit, backgroundcolor = \color{lightgray}, framexleftmargin = 0.2em, framexrightmargin = 1em]
{ 
  "context": "<context>",
  "users": <num_users>,
  "seq_len": <trace sequence length>,
  "normalize": "[pos|minmax]",
  "shuffle": <true|false>
}
\end{lstlisting}
All JSON fields are optional. Context defaults to global context, users to 1, seq\_len to 100, normalize to 
"pos"~\footnote{all values are positive as opposed to forced into 0-1 range as with minmax}, 
shuffle~\footnote{shift values a random timestep forward with values falling off the end inserted at the beginning}
to false.

\subsection*{Response}
If format=json (default). Last dimension in the array is the 2-tuple $\{download,upload\}$.
\lstset{basicstyle=\scriptsize\ttfamily,showspaces=false,showstringspaces=false} 
\begin{lstlisting}[commentstyle=\color{blue}\textit, backgroundcolor = \color{lightgray}, framexleftmargin = 0.2em, framexrightmargin = 1em]
{ 
 "trace": 
    <{user,seq_len,2}-dimensional array>
}
\end{lstlisting}
If format=text
\lstset{basicstyle=\scriptsize\ttfamily,showspaces=false,showstringspaces=false} 
\begin{lstlisting}[commentstyle=\color{blue}\textit, backgroundcolor = \color{lightgray}, framexleftmargin = 0.2em, framexrightmargin = 1em]
<user1 epoch1 download> <user1 epoch1 upload>
<user1 epoch2 download> <user1 epoch2 upload>
...
<user1 epoch<seq_len> download> <user1 epoch<seq_len> upload>

<user2 epoch1 download> <user2 epoch1 upload>
<user2 epoch2 download> <user2 epoch2 upload>
...
<user2 epoch<seq_len> download> <user2 epoch<seq_len> upload>

...
<user<users> epoch1 download> <user<users> epoch1 upload>
<user<users> epoch2 download> <user<users> epoch2 upload>
...
<user<users> epoch<seq_len> download> <user<users> epoch<seq_len> upload>
\end{lstlisting}
Note, an empty line separates two user traces.

\subsection*{Example}
Request
\lstset{basicstyle=\scriptsize\ttfamily,showspaces=false,showstringspaces=false} 
\begin{lstlisting}[commentstyle=\color{blue}\textit, backgroundcolor = \color{lightgray}, framexleftmargin = 0.2em, framexrightmargin = 1em]
POST /generate?format=json

{
 "context":"STREAM_HIGH",
 "seq_len": 3,
 "users": 2
}
\end{lstlisting}
Response
\lstset{basicstyle=\scriptsize\ttfamily,showspaces=false,showstringspaces=false} 
\begin{lstlisting}[commentstyle=\color{blue}\textit, backgroundcolor = \color{lightgray}, framexleftmargin = 0.2em, framexrightmargin = 1em]
{
 "trace":
  [ 
    [0.3,0.2],
    [0.1,0],
    [0.6,0.8]
  ],
  [
    [0.1,0.2],
    [0.5,0.1],
    [0.7,0.3]
  ]
}
\end{lstlisting}

%% file: ns3example.tex
\section{NS3 Example}\label{ns3example}
\lstset{language=C++,basicstyle=\scriptsize\ttfamily,showspaces=false,showstringspaces=false} 
\begin{lstlisting}[commentstyle=\color{blue}\textit, backgroundcolor = \color{lightgray}, framexleftmargin = 0.2em, framexrightmargin = 1em, multicols=2]
#include "mass.h"

...

NodeContainer nodeContainer;
nodeContainer.Create (2);
// Setup of Wi-Fi, LTE or CSMA or other 
// network omitted

...

InternetStackHelper internet;
internet.Install (nodeContainer);
Ipv4AddressHelper ipv4;
ipv4.SetBase ("10.1.1.0", "255.255.255.0");
Ipv4InterfaceContainer interfaceContainer = 
    ipv4.Assign (devices);

double runTime = MassHelper(interfaceContainer,
                            nodeContainer)
      .SetEpochTime(10.5)
      .SetProtocol("tcp")
      .SetMessageSize(2048)
      .SetMaxUpRate(5)
      .SetMaxDownRate(100)   
      .EnableAppContext()
      .SetInitialApp("INTERACT")
      .SetStreamStayProbability(0.8)
      .SetInteractStayProbability(0.6)
      .SetTrace("/data/sample.trace")
      .SetClientIndex(3)
      .SetServerIndex(4)
      .Init();

Simulator::Stop (Seconds (runTime));
Simulator::Run ();
Simulator::Destroy ();
\end{lstlisting}

%% file: android.tex
\section{Android MASS App}\label{androidapp}
\begin{figure}[htpb]
	\centering
	\begin{subfigure}[b]{0.49\textwidth}
		\includegraphics[width=\textwidth]{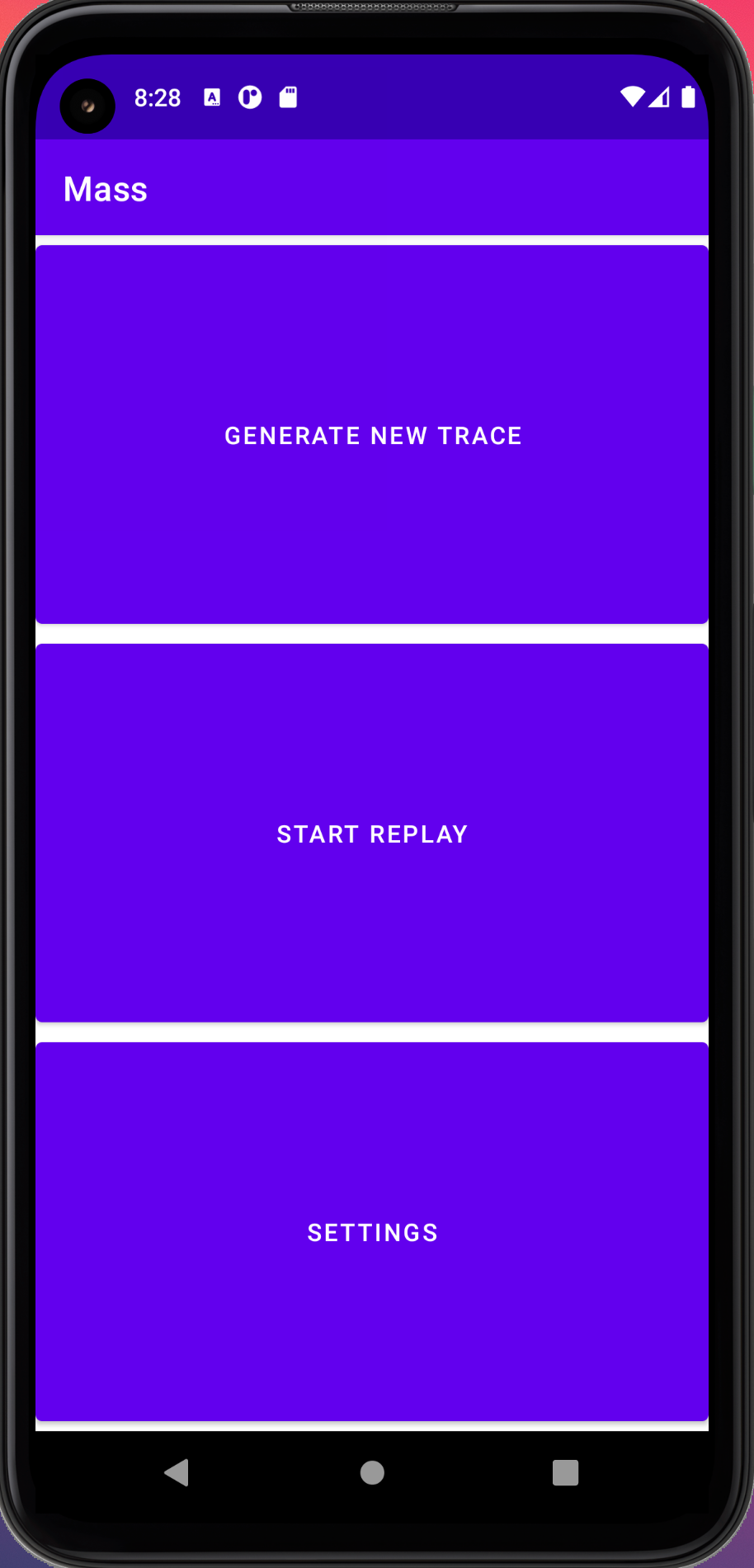}
	\end{subfigure}
	\begin{subfigure}[b]{0.49\textwidth}
		\includegraphics[width=\textwidth]{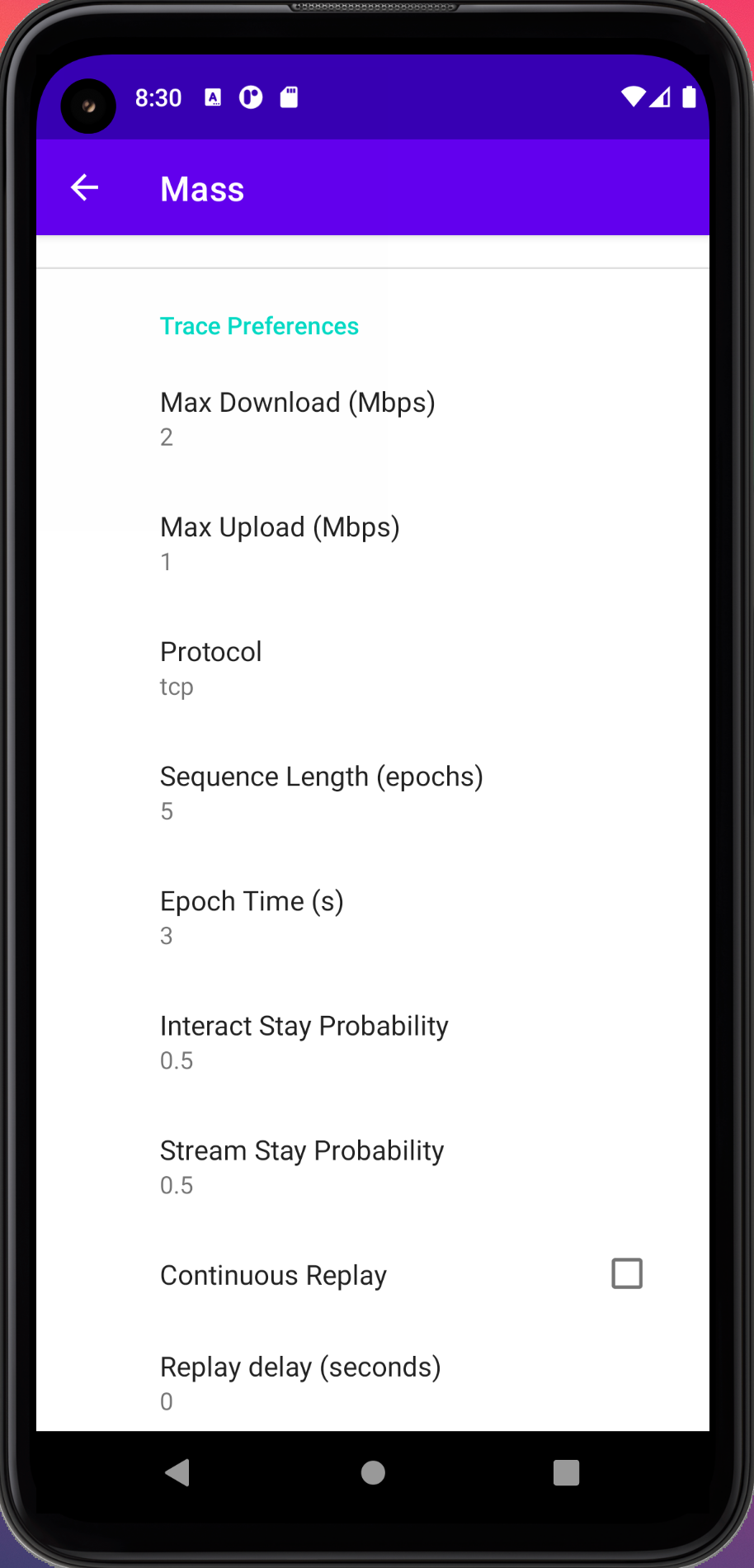}
	\end{subfigure}
	\caption{Android MASS App Screenshots}
	\label{bias}
\end{figure}